\documentclass[final,twocolumn]{elsarticle}

\usepackage{graphicx}

\usepackage{amssymb}

\begin{document}

\begin{frontmatter}



\title{Particle identification using  digital pulse shape discrimination in a nTD silicon detector with a 1 GHz sampling digitizer}


\author[a,b]{K.~Mahata}\ead{kmahata@barc.gov.in}
\author[a,b]{A.~Shrivastava}
\author[a]{J.~A.~Gore}
\author[a,b]{S.~K.~Pandit}
\author[a]{V.~V.~Parkar}
\author[a]{K.~Ramachandran}
\author[a]{A.~Kumar}
\author[a,b]{S.~Gupta}
\author[a]{P.~Patale}
\address[a]{Nuclear Physics Division, Bhabha Atomic Research Centre, Mumbai 400 085, INDIA}
\address[b]{Homi Bhabha National Institute, Anushakti Nagar, Mumbai 400 094, INDIA}

\begin{abstract}
In beam test experiments have been carried out for particle identification using digital pulse shape analysis in a 500~$\mu$m thick Neutron
Transmutation Doped (nTD) silicon detector with an indigenously developed FPGA based 12 bit resolution, 1 GHz sampling digitizer. The nTD Si detector was used in a low-field injection setup to detect light heavy-ions produced in reactions of $\sim$ 5 MeV/A $^{7}$Li and $^{12}$C beams on different targets. 
Pulse height, rise time and current maximum  have been obtained from the digitized charge output of a high bandwidth charge and current sensitive pre-amplifier. 
 Good isotopic separation have 
been achieved using only the digitized charge output in case of light heavy-ions. The setup can be used for charged particle spectroscopy in nuclear reactions
involving light heavy-ions around the Coulomb barrier energies. 
\end{abstract}

\end{frontmatter}

\section{Introduction}
Digital Pulse Shape Analysis  for particle identification  using Si detector is currently under investigation worldwide. 
It has certain advantages over the conventional $\Delta$E-E technique, which is commonly used  for particle identification. 
While the $\Delta$E-E technique requires a thin (a few tens of $\mu$m)  $\Delta$E and a relatively thick (few hundred $\mu$m) E detector, particle identification can be achieved in a single E detector by utilizing Pulse Shape Analysis (PSA) technique. 
Hence, use of PSA technique eliminates the requirement of thin $\Delta$E detectors. Further, the $\Delta$E-E technique can not be applied for low energy particles, which are stopped in the $\Delta$E detector, where PSA technique may still work.
It is proposed to use the PSA technique in most of the future charged particle detector arrays such as FAZIA~\cite{Faz}, GASPARD~\cite{Gas}, TRACE~\cite{Tra} and HYDE~\cite{Hyd}  to reduce the number of channels, cost and complexity. 
  
\begin{figure}[t!]
\includegraphics[width=8cm]{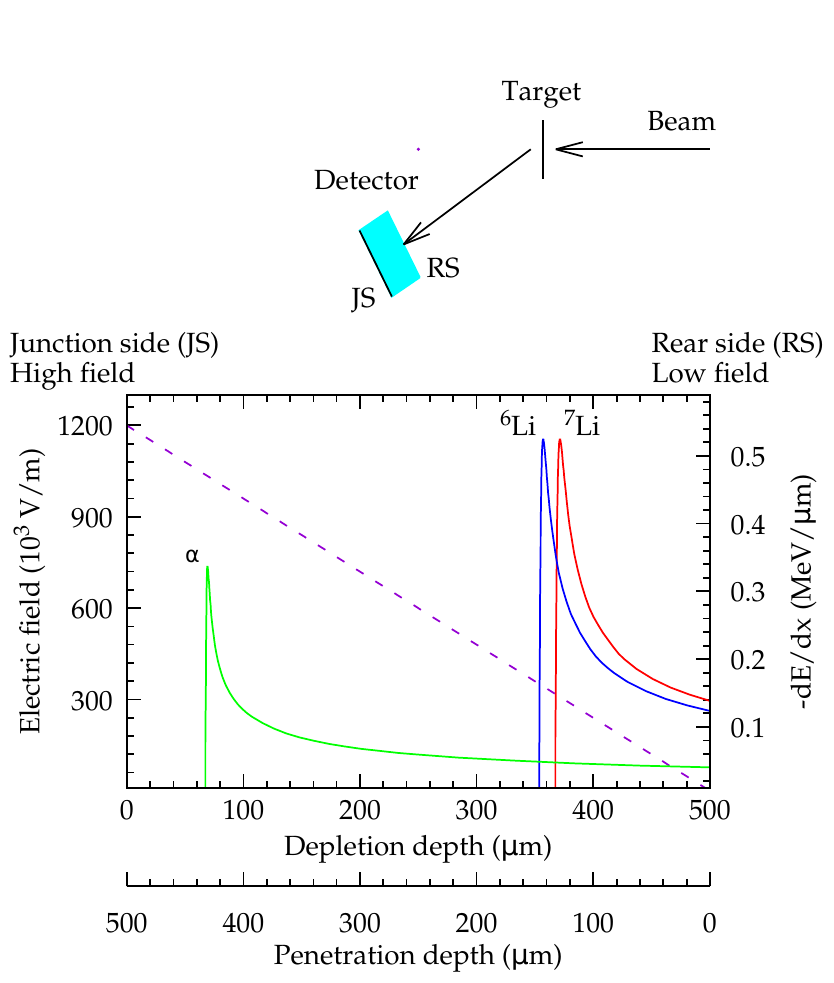}
\caption{\label{range} The schematic diagram of the low-field (rear side) injection setup. The electric field distribution as a function of depletion depth in the 500~$\mu$m Si detector with minimum depletion voltage (225~V) is shown at the bottom with dashed line. The calculated  energy losses~\cite{srim10} as a function of penetration depth for
30 MeV $\alpha$ and $^{6,7}$Li in low-field (rear side) injection mode are also shown.}
\end{figure}

Charged ions, while passing through a medium, lose most of their energy near the end of the track (Bragg peak). As shown in Fig.~\ref{range},
the length of the track (range)  and  the density of ionization/ plasma depend on the charge, mass and energy of the incident ion. The drift of these  produced charges under the applied electric field in a detector results in an electrical pulse. Initially, the plasma produced by the incident ion disturbs the applied electric field locally.
Some time is required to erode the plasma and restore the electric field, which depends on the density of the plasma and the electric field strength~\cite{Bohne85}. 
After the plasma is eroded, the charges drift along the electric field towards the electrodes. The drift time depends on the drift path as well as on the electric field strength. As a result, the pulse shape depends on the charge, mass and energy of the incident ion. Exploiting this fact, separation between deuterons and $\alpha$ particles was first demonstrated by Ammerlaan {\it et al.}~\cite{Ammer63}. If the particles are injected from the rear side, where the electric field is low, the heavier ions will be stopped nearer to the entrance creating denser plasma there as compared to the lighter ions (see Fig.~\ref{range}). As a result both the plasma erosion time and the drift time will be larger for heavier ions as compared to the lighter ions. Thus the rear side (low-field) injection make the PSA technique more sensitive. However, particle identification using PSA requires  Si crystal with highly uniform resistivity and fast pulse processing~\cite{Mutt00}.  

Pioneering work for particle identification in Neutron Transmutation Doped (nTD) Si detector using digital pulse processing has been carried out by the FAZIA collaboration~\cite{Faz}. In a recent work, Pastore {\it et al.}~\cite{Pasto17} have reported  isotopic identification for  Z = 6 - 20 with an energy threshold of 7~MeV/A for oxygen isotopes using specially constructed Si detectors, low noise and high performance digitizing electronics along with optimized PSA algorithms. In their work, only {\it Z} identification was possible for {\it Z} = 3-5. Particle identification for   proton, deuteron, triton  at low energy (3 to 10~MeV) have been successfully demonstrated by Due\~nas {\it et al.}~\cite{Duenas12}. Mengoni {\it et al.}~\cite{Meng14} have also demonstrated identification of protons, deuterons, tritons  and $\alpha$ particles for proton energy down to
2~MeV.

In the present paper, we report the results of particle identification study for light heavy ions  in $^{7}$Li,$^{12}$C + $^{27}$Al, $^{93}$Nb reactions, using pulse shape analysis technique with a nTD Si detector and an indigenously developed FPGA based digitizer. 

\begin{figure}[t!]
\includegraphics[width=8cm]{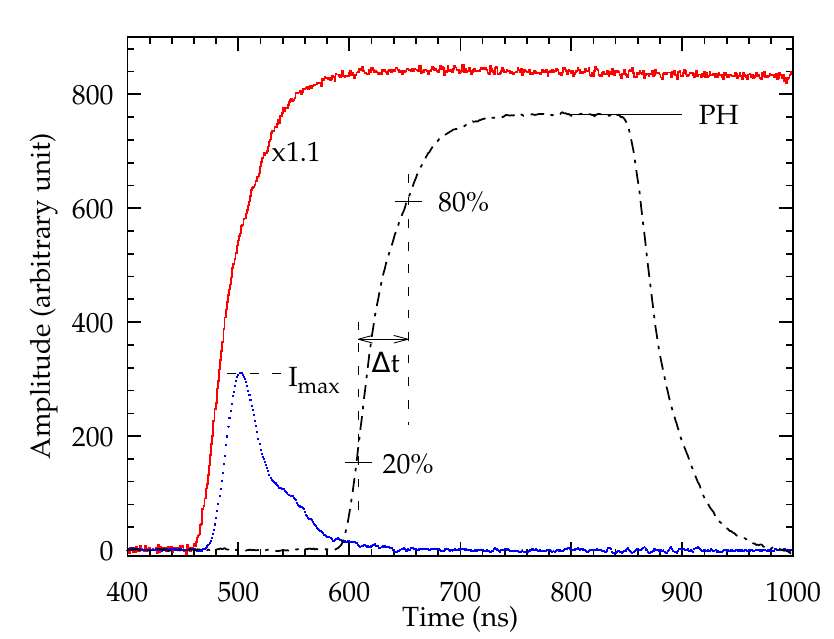}
\caption{\label{pulse} A typical digitized charge  output (multiplied by 1.1 for clarity) of the preamplifier (PACI) using the FPGA based 1 GHz sampling digitizer is shown as continuous red line.  The digitized charge output after passing through the trapezoidal filters (see text) with L = 5~ns, G = 20~ns (blue dotted line) and L = 8~ns, G = 256~ns (black dash-dotted line) are also shown.   The Pulse Height (PH), rise time ($\Delta$t) and current maximum (I$_{max}$) are marked.}
\end{figure}

\section{Experimental Details}
The  experiment was carried out by bombarding $^7$Li and $^{12}$C ions on $^{27}$Al and  $^{93}$Nb  targets at the BARC-TIFR Pelletron LINAC facility, Mumbai. A 500~$\mu$m thick 2~cm~$\times$~2~cm nTD Si pad detector was mounted in the low field  injection mode with a 6~mm diameter collimator on one of the moving arm of the 1.5~m diameter scattering chamber at 30$^\circ$ with respect to the beam direction. The detector is similar to the ones used in Ref.~\cite{Duenas12,Carboni12} and the other details of the detector can be found there. Pulses from the nTD Si detector were processed using a Charge and current (I) sensitive Pre-Amplifier (PACI)~\cite{Hamr04}, which provides a charge as well as a current output. Special care was taken to reduce the noise level. Only the charge output was digitized using an indigenously developed FPGA based digitizer with 1 GHz sampling speed and 12 bit ADC resolution~\cite{Gore12}. Digitized samples were recorded for a time interval  of 2~$\mu$s including a  500~ns pre-trigger interval  with respect to an internal trigger generated by level-crossing. Both the recording interval and the pre-trigger length are programmable up to 32~$\mu$s. Data was also recorded with other time intervals. The digitizer has provision for external trigger also. A typical pulse recorded during the experiment is shown in Fig.~\ref{pulse}.  Effect of bias voltage in the range of 150~V to 300~V was studied to find an optimum operating voltage.

\begin{figure}[t!]
\includegraphics[trim=0.cm 0cm 0cm .0cm, clip=true,width=8cm]{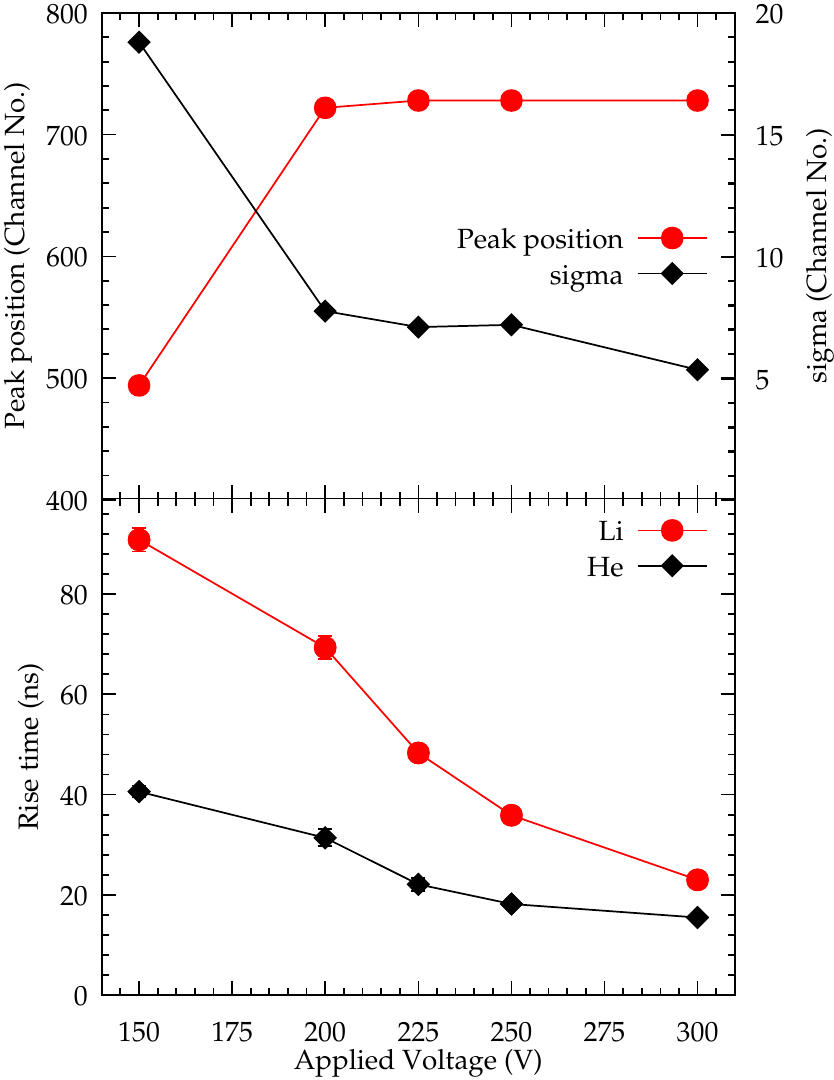}
\caption{\label{peak} (Top) Position and width ($\sigma$) of the elastic peak, (Bottom) rise time for the elastically scattered $^{7}$Li and $\alpha$ particle of similar energy for  $^7$Li + $^{93}$Nb reaction at 35 MeV as a function of applied voltage in low-field injection mode are also shown.}
\end{figure}

\begin{figure}[t!]
\includegraphics[width=8cm]{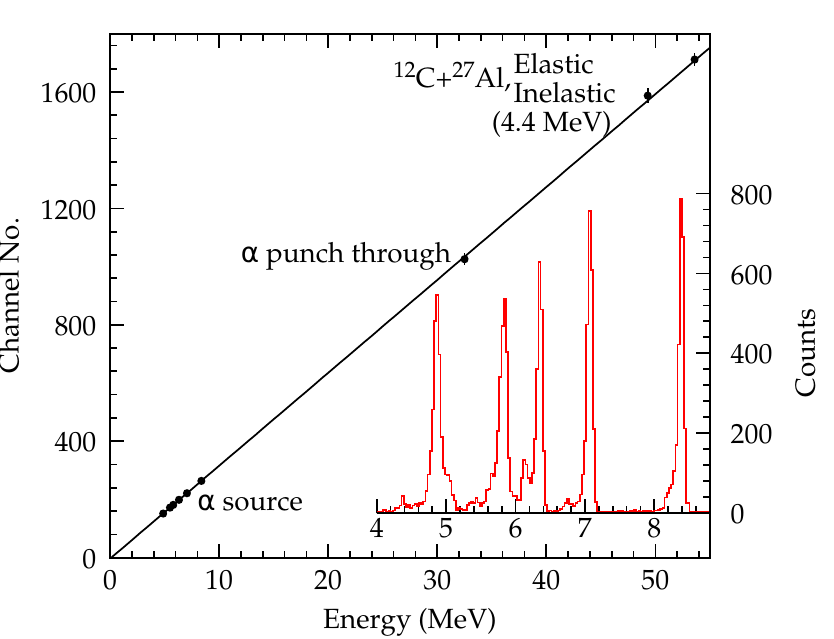}
\caption{\label{calib} Energy vs. pulse height using bias voltage of 225 V for $\alpha$-particles from  $^{229}$Th and $^{241}$Am sources along with $\alpha$-particle punch-through energy, elastic   and inelastic ($^{12}$C, 2$^+$, 4.4 MeV) peak  in $^{12}$C+$^{27}$Al reaction at 63 MeV. The continuous line shows the linear fit. The $\alpha$-particle spectrum for the $^{229}$Th source is also shown in the inset.}
\end{figure}

\section{Data analysis and results}
The digitized pulses were analyzed  using ROOT analysis framework~\cite{root}. Trapezoidal filters were applied to the recorded pulses to obtain the pulse height and the rise time information. The length of the averaging interval (L) of the trapezoidal filter was varied from 8~ns  to 512~ns. The gap (G) between the averaging intervals was kept at 256~ns. Height of the pulse after passing through the trapezoidal filter was taken as the energy of the detected particle and the rise time  of the pulse was taken as the time difference between the 20\% and 80\% of maximum  pulse height. The pulse height was found to be not very sensitive to the length of the averaging interval. An averaging  length of 8~ns, which gives better separation for Li isotopes, have been used in the present analysis. A typical digitized signal after passing through the trapezoidal filter with L = 8~ns  and G = 256~ns is shown in Fig.~\ref{pulse}.  Increasing the length of the averaging interval (L) was found to improve the particle identification for heavier ions with larger rise time.  

As shown in Fig.~\ref{peak}, the energy and time resolution were found to improve with increasing bias voltage. However, the differences in rise time for different type of particles were found to decrease with increasing bias voltage. The optimum voltage was found to be 225 V, which is  also the minimum full-depletion voltage.  At 225~V bias voltage, the energy and time resolutions ($\sigma$) were 300 keV and 1.2~ns for the elastic peak of 35 MeV $^7$Li on $^{197}$Au target. Apart from the detector resolution the width of the
elastic peak has contributions from energy spread due to non uniformity in target thickness, finite angular opening of the detector and inelastic excitations. The width of the 8.4~MeV $\alpha$ peak of $^{232}$Th source was  35 keV at the same bias voltage.

Since the detector is being operated at just the minimum full-depletion voltage (225 V), lower than the voltage at which this type of detectors were operated earlier~\cite{Duenas12,Carboni12} and being used in the low-field injection mode, it is important to check the linearity in the pulse height response. The energy calibration (see Fig.~\ref{calib}) has been performed using 
$\alpha$-particles energies from radioactive sources ($^{241}$Am, $^{229}$Th), $\alpha$-particle punch-through energy and scattered energies of $^{12}$C on $^{27}$Al target. The energy response was found to be linear for the entire energy range covered in the present study and independent of the particle type.

\begin{figure}[t!]
\includegraphics[trim=0.cm 0cm 0cm .0cm, clip=true,width=8cm]{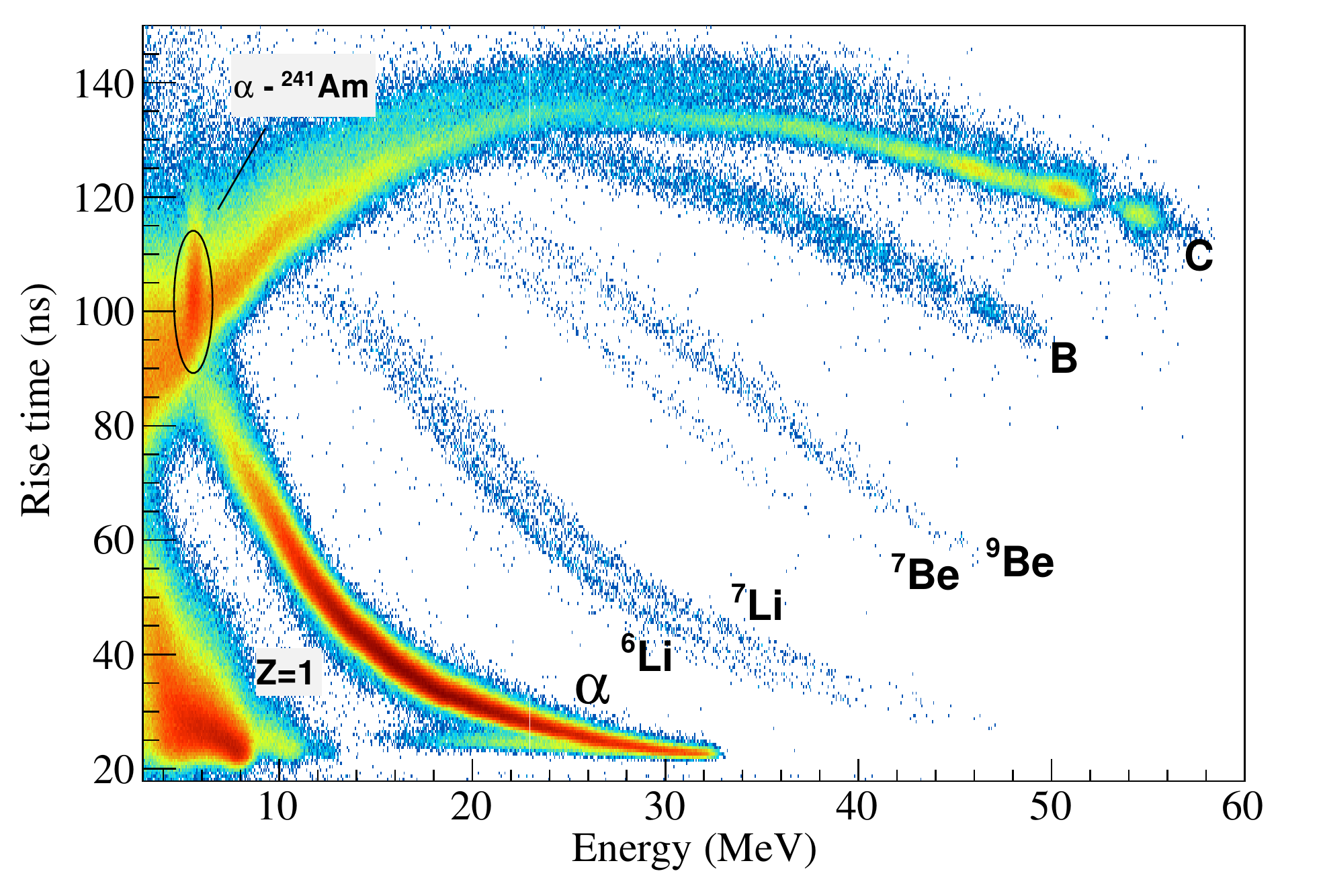}
\caption{\label{CAl} Detected energy vs. rise time plot for the different reaction products in $^{12}$C+$^{27}$Al reaction at 63 MeV. Bands corresponding to different particles are marked. The peak corresponding to the $\alpha$ particles from the $^{241}$Am source, which was in the line of sight of the detector, is also marked.}
\end{figure}

Fig.~\ref{CAl} shows a two dimensional plot of the detected energy vs. rise time for the reaction products detected using the nTD detector placed at 30$^\circ$ from the reaction of 63 MeV $^{12}$C ions with $^{27}$Al target. 
Different bands in Fig.~\ref{CAl} correspond to different type of particles. As energy increases range also increases and the particles deposit energy more close to the junction (high-field region) in a low-field injection setup, decreasing  rise time for a given type of particle. High energy particles  with range larger than the thickness of the detector will not be  stopped and the energy loss in the detector will decrease with increasing energy. This results in back bending as observed in the $\alpha$ band around 32.5~MeV. The average energy at which this back bending occurs is referred as ``punch-through energy". At very low energy, different bands merge together impeding particle identification. At these energies (E/A $\sim$ 1 MeV) the range of the ions are dominated by straggling and consequently the pulse shape become independent of the ion charge and mass.
As can be seen in Fig~\ref{CAl}, good isotopic separation have been achieved in case of Li and Be isotopes. The particle identification obtained using PSA technique was found be comparable with the $\Delta E$-E technique. 
The lowest band in the Fig.~\ref{CAl} corresponds to isotopes of Z=1. Punch-throughs for proton, deuteron and triton can be seen at 8, 10.6 and 12.6~MeV, respectively. However, clear isotopic identification was not possible for Z=1 band. 
For the heavier ions (B,C), different isotopes were not populated with sufficient intensities in present reaction.

\begin{figure}[t!]
\includegraphics[trim=0.cm 0cm 0cm .0cm, clip=true,width=8cm]{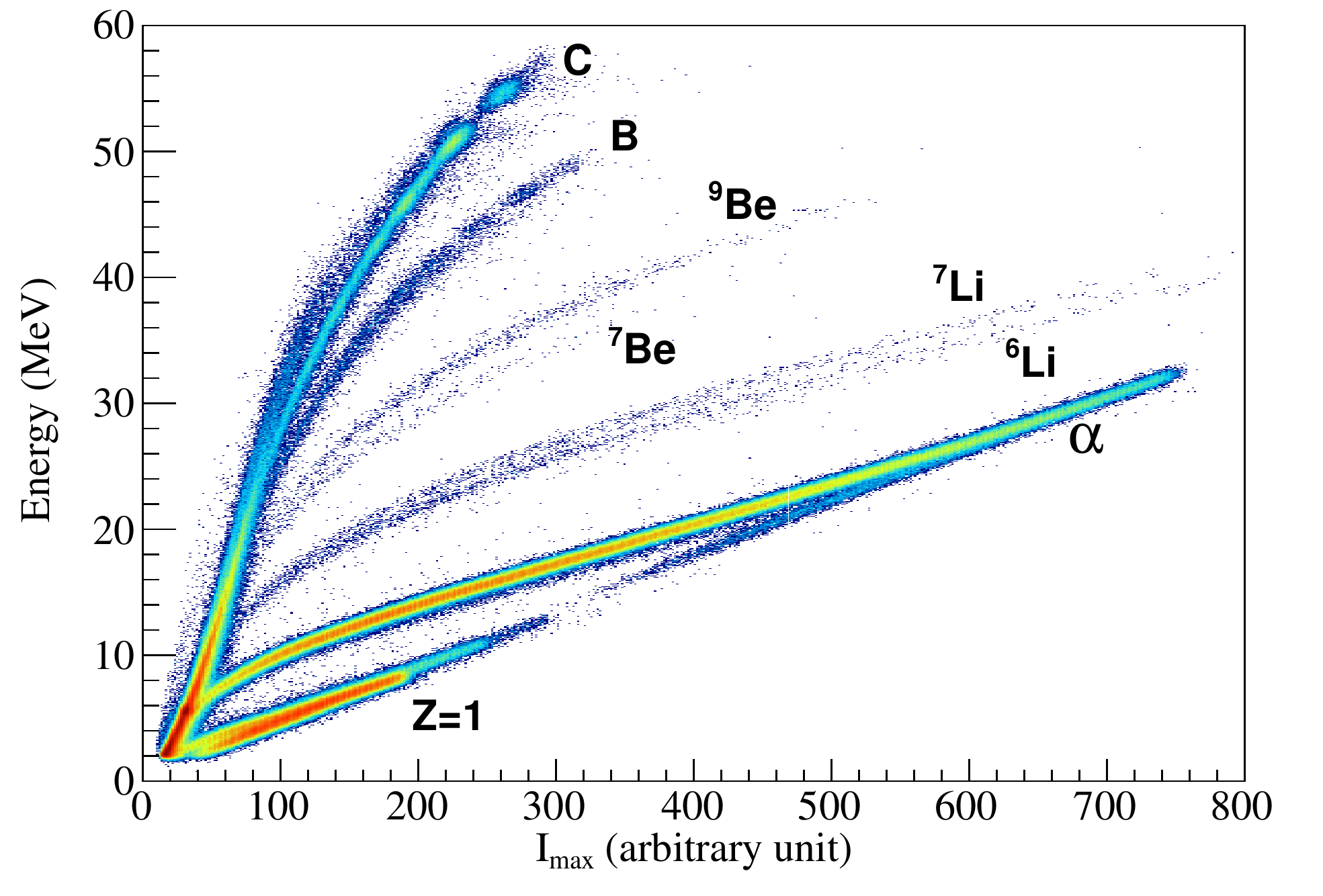}
\caption{\label{current} Energy vs. current maximum (I$_{max}$) plot for the different reaction products in $^{12}$C+$^{27}$Al reaction at 63 MeV. Bands corresponding to different particles are marked.}
\end{figure}

\begin{figure}[t!]
\includegraphics[trim=0.cm 0cm 0cm .0cm, clip=true,width=8cm]{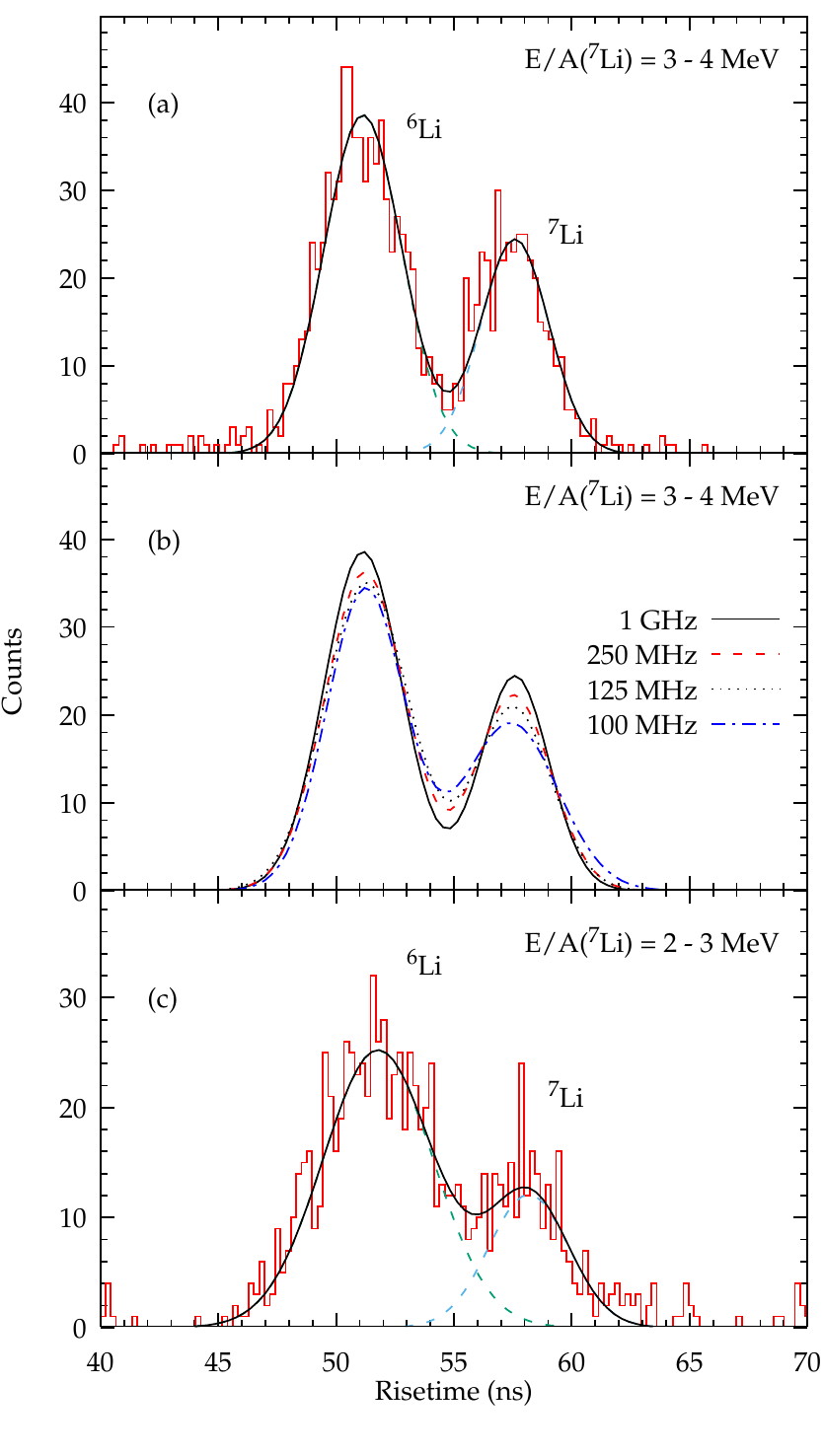}
\caption{\label{Li}  (a) Rise time distributions, corrected for rise time variation with energy,  for Li isotopes with E/A($^7$Li)~=~3~-~4~MeV in $^{12}$C+$^{27}$Al reaction using 1~GHz sampling along with two Gaussian fit. (b) Same as (a) for different sampling frequency, only Gaussian fits are shown for better clarity.(c)  Same as (a) with E/A($^7$Li)~=~2~-~3~MeV.  }
\end{figure}

Conventionally, pulse height obtained from the digitized charge output and the maxima of the separately digitized current output is used for particle identification~\cite{Pasto17,Duenas12}. In the present study, only the charge output has been digitized. However, the corresponding current signal can be generated offline from the digitized charge output. Fig.~\ref{pulse} shows the digitized charge output after passing through a trapezoidal filter with L = 5~ns  and G = 20~ns. These parameters are chosen to qualitatively match the derived pulse with the digitized current pulse shown in Ref.~\cite{Pasto17,Duenas12}. The maxima of the derived pulse (I$_{max}$) vs the energy (pulse height) has been plotted in Fig.~\ref{current}. Similar isotopic resolution, as obtained from the rise time determination, has been observed.

In order to quantify the particle identification capability, we have studied the rise times for the $^{6,7}$Li isotopes. As shown in Fig.~\ref{CAl}, the rise time depends on the energy for a given type of particle. This energy dependence is 
corrected by fitting the rise time as a function of energy using a 2$^{nd}$ order polynomial. We have assumed the energy dependence to be same for both the isotopes. Fig.~\ref{Li} (a) shows the rise time distributions, corrected for the energy dependence, for the Li isotopes with E/A($^7$Li) = 3 - 4~MeV using 1~GHz sampling along with two Gaussian fit. We have also simulated the response of lower sampling frequencies by skipping samples of the recorded pulses.  Fig.~\ref{Li} (b) shows the rise time distributions for different sampling frequencies. For better clarity only the fitted distributions are plotted. 
Particle identification capability was found to reduce gradually with reducing sampling frequency. The rise time distribution for Li isotopes for E/A~=~2~-~3~MeV with 1~GHz sampling along with two Gaussian fit is shown in   Fig.~\ref{Li} (c). The mass resolution is found to be poorer for  E/A($^7$Li)~=~2~-~3~MeV than that for  E/A($^7$Li)~$>$~3~MeV
\begin{figure}[t!]
\includegraphics[trim=0.cm 0cm 0cm .0cm, clip=true,width=8cm]{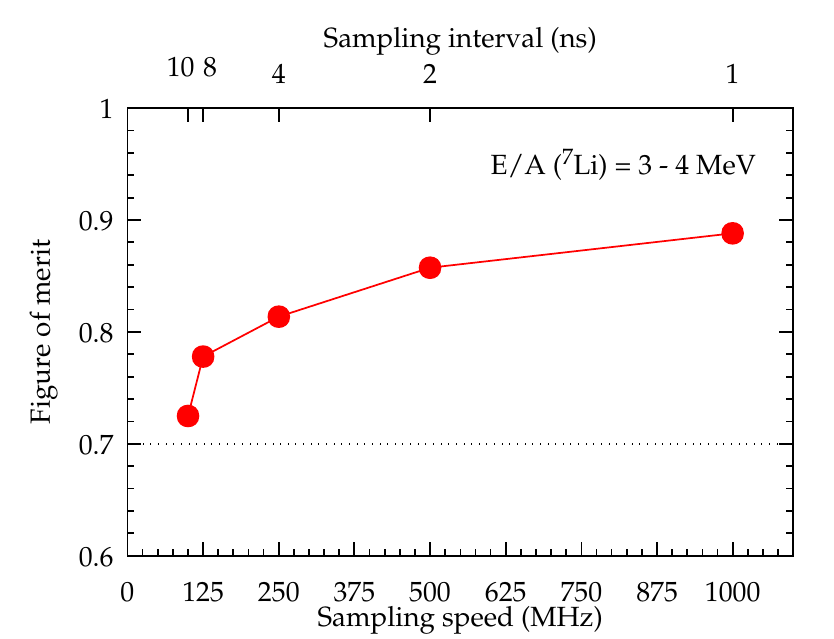}
\caption{\label{fom} Figure-of-merit as function of sampling frequency for Li isotopes with E/A($^7$Li)~=~3 - 4~MeV in $^{12}$C+$^{27}$Al reaction.}
\end{figure}

From the two Gaussian fit, we have evaluated the figure of merit (FoM). FoM is defined as
\begin{equation}
FoM = \frac{Peak_1 - Peak_2}{FWHM_1+FWHM_2}
\end{equation}
where, Peak$_i$ and FWHM$_i$ are the peak position and FWHM of the i$^{th}$ peak. Fig.~\ref{fom} shows the FoM as function of sampling frequency for Li isotopes with E/A($^7$Li)~= 3 - 4~MeV in $^{12}$C+$^{27}$Al reaction. Two peaks are considered to be well resolved if the FoM is larger than 0.7~\cite{Pasto17}. While the FoM is found to gradually reduce with reducing sampling frequency down to 125 MHz, a sharp decrease is observed in going from 125~MHz to 100~MHz. Similar observation was also made by Assi\'e {\it et al.}~\cite{Assie15}.
Recently, it has been reported that the spline and cubic interpolation of the digitized current pulse using a 250 MHz sampling ADC improves the isotopic resolution~\cite{Pasto17}. Spline interpolation of the digitized charge samples did not show any significant improvement in the the isotopic resolution for the sampling frequencies considered in the present study. For a given sampling frequency, the FoM was found to remain constant in the energy range E/A($^7$Li) = 3 - 6~MeV. The FoM for  E/A($^7$Li)~=~2~-~3~MeV with 1~GHz sampling was 0.67, which is slightly below the resolving limit (0.7).

\section{Summary \& Conclusion }
Pulse shape discrimination capability of a nTD Si detector in the low-field injection mode has been studied using an indigenously developed 12~bit resolution, 1~GHz sampling  FPGA based digitizer. Pulse height response of the detector, operated at the minimum full-depletion voltage in the low-field injection mode, is found to be linear and same for different particles detected. Good isotopic resolution using only the digitized charge signal has been demonstrated for Li and Be isotopes for E/A~$>$~3MeV.  Particle identification capability of the nTD Si using PSA technique was found to be comparable to that of the $\Delta$E-E telescope. Present development is aimed to study the pulse shape discrimination capability of double-sided silicon strip detector being developed in India~\cite{Topkar16} for GASPARD collaboration. This technique can be used for charge particle spectroscopy in reaction studies around the Coulomb barrier involving heavy-ion beams.

\section*{Acknowledgment}
Authors are thankful to IPN-Orsay, France for providing the nTD Si detector and the preamplifier (PACI). The authors would like to thank D. Beaumel for careful reading of the manuscript and useful suggestions. The authors are also thankful to the operation staff of the BARC-TIFR Pelletron-Linac facility, Mumbai, for the excellent support during the experiment.

\bibliographystyle{elsarticle-num}
\providecommand{\noopsort}[1]{}\providecommand{\singleletter}[1]{#1}%

\end{document}